\documentclass[12pt,preprint]{aastex}
\usepackage{epstopdf}

\shorttitle{Cosmological Unimportance of LSBs}
\shortauthors{Hayward et al.}

\begin{document}

\title{The Cosmological Unimportance of Low Surface Brightness Galaxies}
\author{C.C. Hayward, J.A. Irwin, J.N. Bregman}
\affil{University of Michigan}
\affil{Astronomy Department, 830 Dennison Building, University of Michigan,
  Ann Arbor, MI 48109-1042}
\email{cchaywar@umich.edu, jairwin@umich.edu, jbregman@umich.edu}

\begin{abstract}
We have searched for Type Ia supernovae (SNe Ia) in the
local ($d \lesssim60$ Mpc) Universe using \textit{Northern Sky Variability
Survey} (NSVS) data collected from the nightly optical
surveys of the \textit{Robotic Optical Transient Search Experiment}
(ROTSE) Telescope.
It was hoped that SNe Ia would provide a means to find 
previously-unknown low surface brightness (LSB) galaxies or
displaced stars that would otherwise be very difficult to detect.
The ROTSE data allowed us to survey 19,000 square degrees at declinations
north of $0^\circ$, but we did not find a single
SN Ia in a period of time covering roughly one year. Using known SNe Ia
rates in bright galaxies, we set an upper limit on the
optical luminosity density, ${\cal L}_B$, of LSBs in the local Universe. Using
mean LSB baryonic and dynamical mass-to-light ratios,
we find 95\% upper limits for LSBs of ${{\cal L}_B} \le 2.53
\times 10^8 L_{B, \odot} \, \hbox{Mpc}^{-3}$, $\Omega_b
\le 0.0040$, and
$\Omega_m \le 0.036$.
We conclude that LSBs and displaced stars are not a major constituent
of matter in the local Universe.
\end{abstract}


\section{INTRODUCTION}

Low surface brightness (LSB) galaxies are usually defined as galaxies with 
B-band central surface brightness $\mu_{0,B} > 23.5$ mag $\hbox{ arcsec}^{-2}$.
The lower central surface brightness of LSB galaxies makes them difficult
to detect against the noise of the night sky. Therefore, they are
systematically absent in many optical galactic surveys. In recent years,
improved techniques to detect LSBs have been developed, and the bulk properties
of LSBs are now better understood \citep[e.g.,][]{bot97,bur01,mon02}. However, the
space density of LSBs is still a matter of debate. It has been suggested that
LSBs could comprise up to 50\% of all galaxies \citep[e.g.,][]{bot97},
and thus could represent a major repository of baryons in the Universe.
If so, we may have to reconsider our notions of galaxy formation,
as the
standard Hubble sequence is solely based on high surface brightness (HSB)
galaxies. Additionally, all
previous galaxy surveys will be considered incomplete, and
the luminosity function of the Universe and large-scale
structure will be affected. Consequently, an important goal is to determine
the contribution of
LSB galaxies to the optical luminosity density ({$\cal L_B$}),
$\Omega_b$,
and $\Omega_m$ of the local Universe.

If LSB galaxies are significant baryonic repositories in the local Universe,
then Type Ia supernovae (SNe Ia) should be detected in LSB galaxies in
addition to HSB galaxies. A SN Ia in a LSB galaxy can have an apparent
magnitude much greater than the magnitude of the host galaxy.
Therefore, such SNe Ia provide a way to find otherwise hard-to-detect LSB
galaxies. Our strategy is to search for SNe Ia in nearby LSB galaxies by
using data from a shallow, all-sky survey (rather than a pencil-beam survey)
that images the sky on a nightly basis.

Note: Results contained here-in are based on the explicit assumption
that the binary frequency in LSB galaxies is the same as in HSB galaxies.

\section{DATA SET}

We used data from the \textit{Northern Sky Variability Survey} (NSVS),
publicly available data \citep{woz04}
collected by the \emph{Robotic Optical Transient Search
Experiment}'s ROTSE-I telescope, an array of four cameras mounted on
a common equatorial platform \citep{keh01}.
The optics are four Cannon 200 mm
focal length, f/1.8, telephoto lenses in FD mounts, which are combined with
four Apogee Instruments AP-10 CCD cameras with Thomson 2048 x 2048 14 $\mu$m
imagers. The pixel width of the CCDs is 14.4'' \citep{ake00}.

Each lens has a 64 square degree field of view, so the array covers
256 square degrees. The ROTSE-I telescope can detect objects brighter than
15.5 mag \citep{keh01}. Images were taken with unfiltered
CCD's, but
magnitude is calibrated to a modified B-band. Magnitude in the ROTSE
band \citep{ake00} is

\begin{equation}
\label{eq:rotse}
m_{\hbox{ROTSE}} = m_V - \frac{m_B - m_V}{1.875}.
\end{equation}

The main purpose of ROTSE is to search for an optical
counterpart to gamma-ray bursters (GRBs). When a GRB is detected, the ROTSE
telescopes quickly move and take an image at the position of the GRB.
When not imaging GRBs, the ROTSE-I telescope was used to take two
images in rapid succession of every visible field each night. These
skypatrols cover the sky north of $-38^{\circ}$ declination twice nightly,
but only the data north of $0^{\circ}$ declination was suitable for our purposes.
The data set we used spans a roughly one year period, from May 2000 to
April 2001, though the length of time covered varies by field.

\section{DATA ANALYSIS}

The ROTSE-I data are useful for detection of Type Ia supernovae
because of the nightly frequency of the skypatrols.
We designed an algorithm
to search for SNe Ia in the ROTSE skypatrols data by analyzing the shape of
the light curve of each of the over 12 million objects in the data set. The
algorithm can detect SNe Ia that peak at $V$ magnitude
14.5 or brighter. Since SNe Ia have a standard absolute $V$ magnitude of 
approximately -19.45
\citep{rie99}, the upper limit of 14.5 (minus absorption) defines the
volume scanned for each field.

Our search is sensitive to Type Ia supernovae with LSB hosts or without hosts altogether.
A SN Ia in an LSB galaxy will have a template light curve shape because the host
galaxy's flux is at least a factor of 10 less than the flux of the supernova.
If the galaxy contribution were brighter, the resulting light
curve would significantly deviate from the SN Ia template curve and the
supernova would not be detected by our algorithm. This is not the case for HSB galaxies,
for which light from a supernova would be blended together with a comparable amount of
galaxy light due to the large pixel size ($14\farcs4$) of the detector, unless the supernova were
well-separated from the galaxy core. Using only fields of low absorption ($A_V \lesssim 0.5$),
distances out to $\sim 60$ Mpc can be searched. Contamination from types of SNe other than Ia
is not significant because the other types peak at a considerably dimmer magnitude than SNe
Ia and therefore would not appear frequently in the volume probed.

\subsection{The Automated Search Algorithm}

In order to find supernovae in a field, we analyzed the light curves
of each object in the field with an automatic filtering program.
The development of this program was based on testing
with Monte Carlo simulations. The filter was designed to maximize
the number of potentially valid supernovae identified and minimize the
number of false positives.

The program is composed of the following steps:

1. The image is trimmed to make the field square and to include only areas of
the sky that were well-observed by ROTSE-I. This step ensures that each area
of the sky is only covered
once. Also, individual observations with magnitudes outside the
calibrated range, 9.0 to 15.5 magnitude, are discarded.

2. Objects for which there are fewer than 10 observations are discarded
because these light curves are too poorly sampled to determine 
whether or not they are supernovae.

3. We remove outliers from the light curves by comparing the light curves to
actual SNe Ia light curves given in \citet{bra98}. This step
eliminates data points that are most likely due to incorrect calibration.

4. Subsequently stricter criteria are applied to the objects' light curves:
we discard observations that
are $> 0.3$ magnitude dimmer than the previous observation if the observations
are
before the maximum and observations that are $> 0.3$ magnitude brighter than
the previous observation if the observations are after
maximum (i.e., we eliminate observations that drop too much when the light
curve is increasing and rise too much when the light curve is decreasing).
Additionally, the maximum and minimum must occur at least five
days apart. These steps further refine the light curves by removing points that
differ from the SN Ia light curve template.

5. Points that differ from the entire light curve's mean by more than
3$\sigma$ are removed. This step should not affect SNe Ia light curves because these
curves have relatively large $\sigma$.

6. If the modified light curve has less than 10 observations, the
object is discarded.

7. A series of filters is applied to the modified light curve. These filters
require the curve to vary with time in a manner consistent with typical SNe
Ia light curves, taken from \citet{bra98}. Specifically, there must be at
least one magnitude variation over the entire curve. We require all objects
with light curves that peak below 12th
magnitude to be visible for less than 150 days. All objects that peak
between 12th and 9th magnitude must show at least 2 magnitudes variation.
These filters eliminate most long-term variable stars.

\subsection{Manual Examination of Results of Filtering Program}

The filters described above yielded a set of possible SNe Ia for each field,
which were subsequently examined by eye.
The number of possible SNe Ia per field ranged from zero to
hundreds, varying directly with Galactic latitude
and inversely with the time coverage of the field.

SIMBAD was used to determine if there are any known objects within
the error box (15'') of the position of each possible SN Ia, in which case,
the object is discarded.
Otherwise, the Digitized Sky Survey (DSS) was used to obtain images of
the sky near the possible SN Ia position. If there is an object visible in
the images at the position of the possible SN Ia, it is unlikely that the
object is an SN Ia in an LSB galaxy, since no supernovae that appeared during
the ROTSE epoch would also appear during the epochs of the DSS images.

\subsection{Testing the Search Algorithm}

In order to test the efficacy of the filter, we generated a set of
$10^3$ synthetic SNe Ia light curves for each ROTSE field. The synthetic
light curves have observations only for days that the actual field
was observed, and the errors of a synthetic observation are randomly
selected from the errors of actual observations with similar magnitude.
The distribution of peak apparent magnitudes of the simulated SNe Ia is
determined by assuming that all the simulated supernovae peak at absolute
magnitude -19.45 and are distributed randomly throughout the field volume.

\begin{figure}
\plotone{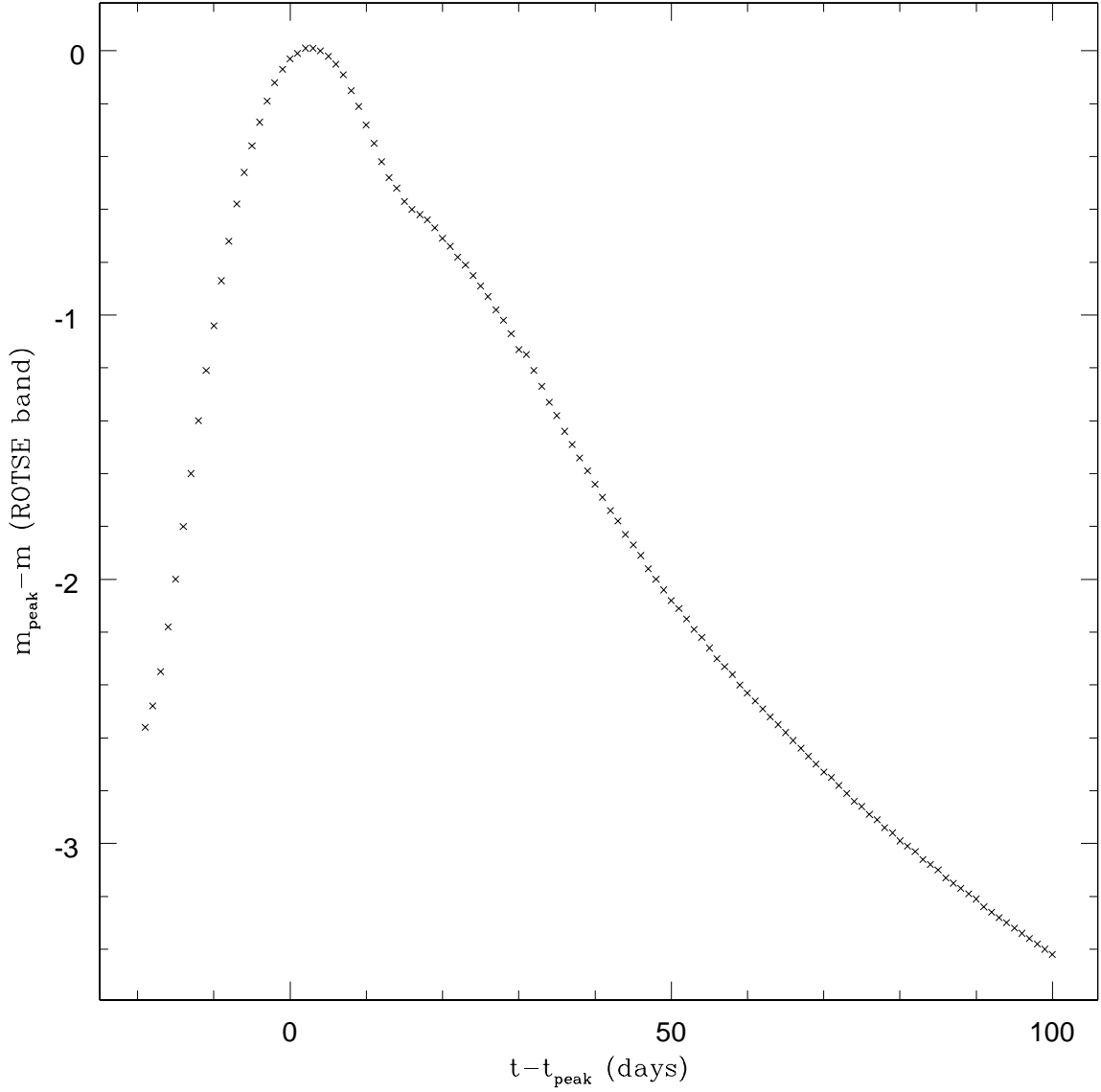}
\caption{SN Ia light curve template in the ROTSE band, created from the
  B-band data of \citet{gol01} and the V-band data of \citet{ham96}.}
\label{fig:template}
\end{figure}

\begin{figure}
\plotone{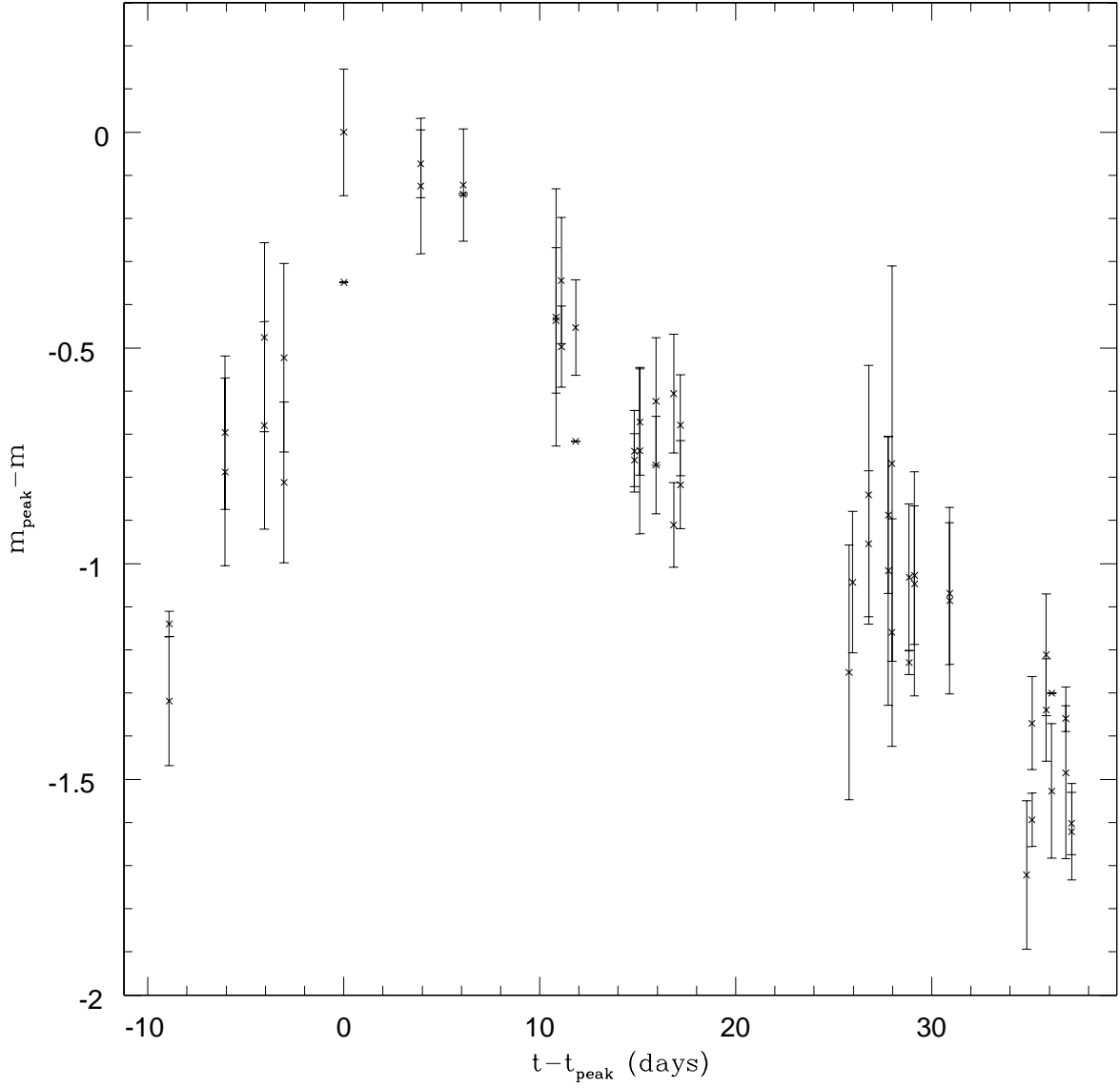}
\caption{Light curve of a simulated SN Ia with peak magnitude 13.59
  generated by our Monte Carlo program. Note that there are two separate
  observations for each day there is data.}
\label{fig:simulated}
\end{figure}

A SN Ia template light curve in the ROTSE
band, created using V-band data from Hamuy et al.\ (1996), B-band
data from \citet{gol01}, and Equation \ref{eq:rotse},
is shown in Figure \ref{fig:template}.
Monte Carlo simulations were used to generate SNe Ia light curves
from this template. An example of a SN Ia light curve (peak magnitude 13.59)
generated by the Monte Carlo simulation program is shown in Figure
\ref{fig:simulated}.

We ran our filter program on each set of simulated SNe Ia, examined the
results by eye, and recorded the percent identified as possible SNe Ia
as the detection rate. For each field, the detection rate
was multiplied by the total volume of space scanned and the total
time the field was observed to obtain the effective detection region
(i.e., the spatial volume and time period in which 100\% of the SNe Ia
peaking in the region would be detected). Dividing the sum of the effective
detection regions for all the fields by the number of SNe Ia found yields
an upper limit on the SNe Ia (SNe Mpc$^{-3}$ century$^{-1}$) rate for LSB
galaxies.

\begin{figure}
\plotone{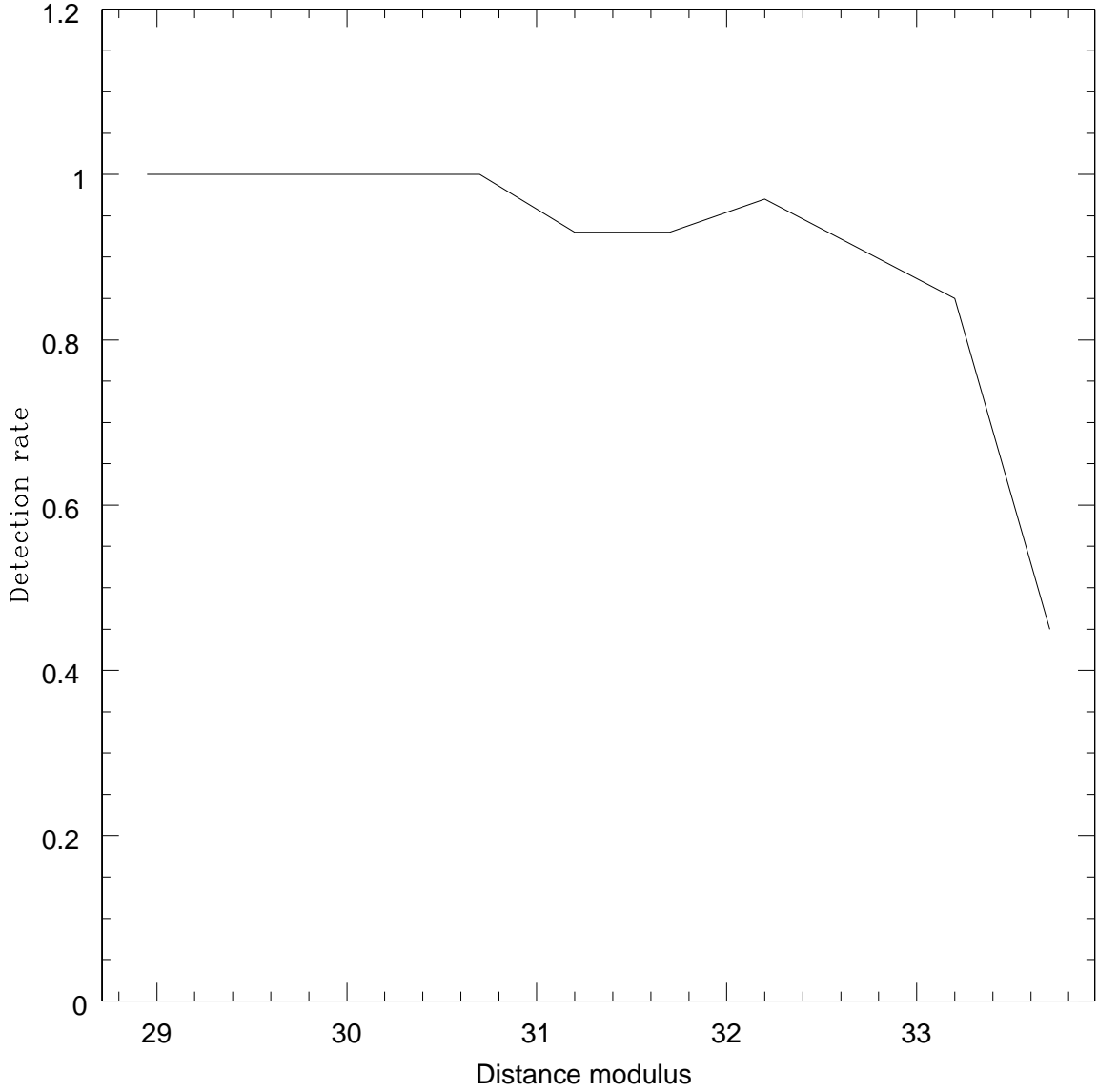}
\caption{Detection rate of the filter vs. distance modulus, binned by 0.5 mag.
  The detection rate is almost 100\% for SNe Ia observed at distance modulus
  less than 32, but the detection rate drops significantly for larger values of the
  distance modulus.}
\label{fig:rate}
\end{figure}

The Monte Carlo simulations determined that the filtering program had a
detection rate of 50-90\% for most fields. The rate was nearly
100\% for simulated SNe Ia peaking at 12th magnitude or higher but was
significantly less for SNe Ia peaking below 12th magnitude. A plot of
detection rate versus magnitude for a field with total detection rate of 66\%
is given in Figure \ref{fig:rate}.

\begin{figure}
\plotone{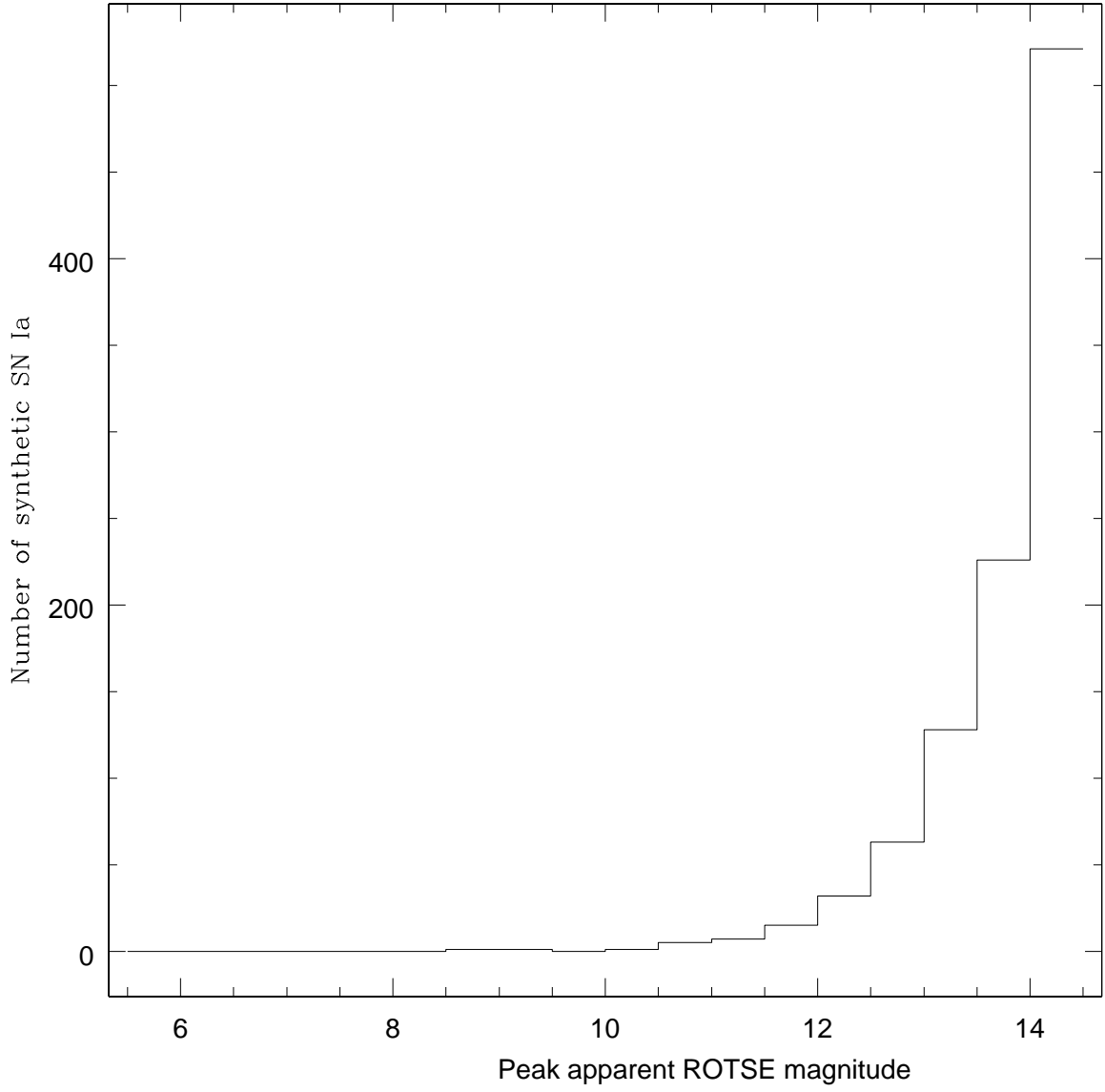}
\caption{This example of the number of SN Ia generated by our Monte Carlo
  simulations as a function of magnitude illustrates how SN Ia are
  preferentially observed at fainter magnitudes.}
\label{fig:occurrence}
\end{figure}

The detection rate is intrinsically tied to the distribution of SNe Ia with
respect to magnitude.
An example of this distribution for $10^3$ SNe Ia generated by the
Monte Carlo simulation is given in Figure \ref{fig:occurrence}. In the
simulation, all SNe Ia peak at an absolute magnitude of -19.45, so the
apparent magnitude of a supernova immediately yields the distance to the
supernova. The probability of detection of a SN Ia is directly
proportional to volume scanned, so SNe Ia are
preferentially observed at fainter magnitudes. Accordingly, the total 
detection rate is dominated by the lower detection rates for
SNe Ia that peak below 12th magnitude.

\subsection{Known SNe Ia in the Data Set}

We tested the filtering program to determine if it could identify any
previously known SNe Ia in the data set. The International Astronomical
Union's List of Supernovae was searched for all SNe Ia that meet the following
requirements: The SN Ia was observable during the time period and area of sky
for which we have data (roughly from May 2000 to April 2001 and north of
0$^{\circ}$ declination); it was bright enough (magnitude
$> 15.5$) to be detected by ROTSE-I; and it was located further than $\sim 30$
arcseconds (2 pixels) from the host galaxy's center. If the SN Ia was too
close to the galaxy, the light curve will not resemble the SN Ia template,
and our filtering program is not designed to detect such objects.

The only SN Ia found that met all three criteria was SN 2001V, which
reached maximum light on March 5, 2001, with B magnitude 14.64 \citep{man01}.
SN 2001V was located 64.6 arcseconds ($\sim 4.5$ pixel widths)
from its host galaxy, NGC 3987, a Type Sb spiral with B-band magnitude
14.4 \citep{vin03}. The position of SN 2001V was imaged frequently
by ROTSE-I
in late February and early March, so the SN Ia should be observed
in the data, though only for a short time.

The filtering program did not identify a possible SN Ia at the position of
SN 2001V. This is expected because the filter can only detect supernovae
that peak at magnitude 14.5 or brighter. SN 2001V peaked at 14.64, so it is
not possible to detect this supernova with the filtering program.

We checked to see if SN 2001V is in the original data set by searching for
all objects observed by ROTSE-I near the position of SN 2001V. A few objects were
found within an arcminute of SN 2001V, but they are not likely to be SN 2001V
because they are a few magnitudes brighter than the supernova and show little
variation. SN 2001V was brighter than magnitude 15.5 for only $\sim 20$ days
\citep{vin03}, so the supernova may not be in the original data set.
Thus the check for known SNe Ia is not possible, as no good candidates exist.

\section{CALCULATIONS AND RESULTS}

The search algorithm yielded many objects for subsequent analysis, but none
are confirmed SNe Ia. Many of the objects are known variable stars, and the
others correspond to an object visible on the DSS. Therefore, we conclude that
each candidate is a variable star.

The search program is effective because many known (and possibly many
unknown) variable stars were found, and some of these objects have light
curves that very closely resemble the SN Ia template.
The Monte Carlo simulations indicate that over half of the
simulated SN
Ia peaking during observation of a field were detected. Since the filter
is effective and detection rates were calculated, finding zero SNe Ia
outside HSB galaxies is significant and can be used to place limits on the
prevalence of LSB galaxies and displaced stars.

Upper limits on the number of SNe Ia in LSB galaxies in the data set were
determined using Poisson statistics. Since we observed no SNe Ia, it was
necessary to use single-sided statistics. There is a 16\% chance
of observing 0
SNe Ia in LSB galaxies if the expected value is 1.84 SNe Ia, so $\le 1.84$ SNe
Ia is the single-sided $1\sigma$ confidence limit. Similarly, there
is a 5\% chance of observing
0 SNe Ia in LSB galaxies if the expected value is 3.00, so the 95\% confidence
limit is 3.00 SNe Ia.

For each field, the product of the detection rate (from the Monte Carlo
simulations)
and the total space and time searched yields an effective
detection region. Summing this over all fields gives the total effective
detection region ($3.33 \times 10^7$ Mpc$^3$ days).

We assume that SNe Ia occur at the same rate (number of SNe Ia per $L_{B,\odot}$)
in LSBs as is observed in the local Universe. One objection to this assumption is that
LSB galaxies may have binary fractions
and/or star formation rates that differ significantly from those of HSB galaxies.
However, binary formation is
a local process \citep{toh02}, so, even though LSBs may have lower star
formation rates than HSBs
\citep{imp97}, the low stellar density of LSB galaxies will not affect
the binary fraction. Though it was originally thought that LSBs are generally
bluer than HSBs \citep{imp97}, the CCD survey of \citet{one97}
found LSBs with colors ranging from very blue to very red, suggesting that
the previous lack of red LSBs was due to a selection effect. Thus the stellar
population of LSBs is likely similar to that of HSBs, which also suggests that the
SN Ia rates are similar. Furthermore, the HSB Type Ia supernova rate can be
divided into a contribution from old progenitors and a contribution from young
progenitors \citep{man05}. The old progenitor contribution is independent of
the star formation rate and accounts for most of the SNe Ia in ellipticals, 50 percent
in S0a/b, 20 percent in Sbc/d, and a few percent in irregulars. This contribution should
also be present in LSBs. The young progenitor contribution is proportional to the
SFR, and thus may be less important in LSBs. Therefore the SN Ia rate may be less
in LSBs than HSBs, but not significantly less in ellipticals and early-type spirals.
Though the assumption that the SN Ia rate in LSBs is the same as that in HSBs is
reasonable, it is still an assumption, and it may be modified as supernovae in LSBs
are studied further.

\citet{cap99} determine that the SN Ia rate in HSB galaxies is
$0.18 \pm 0.05$ SNe century$^{-1} 10^{-10} L_{B,\odot}$, so a lower limit on the SN
Ia rate of $\ge 0.13$ SNe century$^{-1}$ $10^{-10} L_{B,\odot}$ is adopted. Dividing
the upper limits on the SN Ia rate per volume in LSB galaxies by the lower
limit on the SN Ia rate per mass \citep{cap99} yields upper
limits on the contribution of LSBs to the optical luminosity density of
the local Universe.

We have calculated a mean LSB baryonic mass-to-light ratio of 2.20 (solar units) from
the results of \citet{deb96}, \citet{mat01}, \citet{mcg97}, \citet{mon03a,mon03b},
\citet{one00}, and \citet{zav03}. From the data of \citet{deb96}, \citet{mcg98}, and
\citet{zav03}, we calculate a mean LSB dynamical mass-to-light ratio of 20.21.
Note that the dynamical ratio is less certain than the baryonic ratio because the number
of LSBs with calculated baryonic mass-to-light ratios is much greater than the number
of LSBs with calculated dynamical mass-to-light ratios. We have used these typical
mass-to-light ratios to set upper limits on the contribution of LSBs to $\Omega_b$
and $\Omega_m$. The results of the calculations are listed in Table~\ref{tab:limits}.

The optical luminosity density, $\cal L_B$, of HSB galaxies is
$(1.35 \pm 0.14) \times
10^8 \, L_{B,\odot} \, \hbox{Mpc}^{-3}$ \citep{fuk04}.\footnote{$H_0 = 71$
\hbox{km} s$^{-1}$ \hbox{Mpc}$^{-1}$ \citep{ben03} is used.}
The $1\sigma$ and 95\% confidence upper limits on the optical luminosity density
of LSBs are slightly above the HSB value, so LSBs and displaced stars
contribute no more than an amount comparable to HSBs to the optical
luminosity density of the local Universe.

We set upper limits on the contribution of LSBs to the mean
$\Omega_b$ and $\Omega_m$ of the local
Universe (Table \ref{tab:limits}).
The upper limit on $\Omega_b$ for LSBs indicates that LSBs
cannot increase the currently observed mean value of
$\Omega_b$ in the Universe by more than 9 percent
because $\Omega_b = 0.044 \pm 0.004$ \citep{ben03}.
Therefore, LSBs are not significant baryonic repositories in the local
Universe. LSBs cannot increase the value of the mean
$\Omega_m$ in the Universe by
more than 13 percent because $\Omega_m = 0.27 \pm 0.04$ \citep{ben03}.
Thus LSBs cannot account for a significant component of the
total dark matter in the Universe.

\section{DISCUSSION AND CONCLUSIONS}

No supernovae were found in LSB galaxies (or outside known galaxies), and
this result was used to set an upper limit on the optical luminosity
density of LSBs and displaced stars in the local Universe. This also
constrains
the contribution of LSBs to $\Omega_b$ and
$\Omega_m$ to be less than 9\% and 13\%, respectively.
Therefore, LSBs do not account for a significant amount of missing baryons
or dark matter.

\citet{spr97} conducted an optical survey using the Advanced Plate
Measuring (APM) system at Cambridge and calculated the luminosity function for
LSB galaxies
in the survey. They found that LSBs contribute approximately 30\% to
the field galaxy luminosity density. This agrees with our result that LSBs
can contribute roughly no more than HSBs to the luminosity density.

\citet{zwa01} conducted a blind strip survey of
HI-selected galaxies in the 21-cm line. Since the survey was HI-selected,
it avoided optical selection effects but only included gas-rich galaxies.
They found no LSBs with central surface brightness $\mu_{0,B} > 24$ mag
arcsec$^{-2}$.
\citet{zwa01} concluded that gas-rich LSBs (defined as $\mu_{0,B} > 23.0$
mag arcsec$^{-2}$) contribute no more than 5 $\pm$ 2\% to the optical
luminosity of the universe and no more than 11\% to $\Omega_m$.

\citet{dri99} conducted a volume-limited sample of 47 galaxies from the
Hubble Deep Field to find the local $(0.3 < z < 0.5)$ bivariate
brightness distribution. Driver\ (1999) found that LSBs
($21.7 < \mu_{0,B} < 24.55$) contribute 7 $\pm$ 4\% to the optical
luminosity density and 12 $\pm$ 6\% to $\Omega_m$.
Note that our upper limit on $\Omega_m$ is in good agreement with both
\citet{zwa01} and \citet{dri99}.

Studies of intracluster starlight provide an indirect probe of the
contribution of LSBs and displaced stars to the optical luminosity
density of clusters. \citet{fel98} conducted a survey for
intracluster planetary nebulae (PN) in the Virgo Cluster. They detected
PN by comparing sums of on-band images to sums of off-band images and
picking point sources that were only visible in [O III]. \citet{fel98}
found that intracluster starlight accounts for between 22 and 61 percent
of the optical luminosity density of the cluster. This result agrees
with what has been found in the local Universe even though tidal disruption
is more common in clusters than in the field.

Various authors have searched for LSBs in the Coma cluster
\citep[e.g.,][]{and02,ulm96,kar95}.
They have found numerous LSBs in the core of the cluster, but the LSBs
are only a minor fraction of the luminosity density of the cluster.
\citet{ber95} did a deep CCD integration of a region in the core
of the Coma cluster and determined that LSBs contribute about
$1/400$ of the luminosity of the cluster.

It is encouraging that numerous different methods of studying the
contributions of LSBs to the optical luminosity density and
$\Omega_m$ yield consistent results, both 
for the local Universe and for clusters. These results all point to
the conclusion that LSBs do not account for a significant amount
of matter.

\acknowledgements

We would like to thank the ROTSE team, with special thanks to Tim McKay and
Carl Akerlof, for allowing us access to their
skypatrol data. ROTSE is a collaboration of Lawrence Livermore National Lab,
Los Alamos National Lab, and the University of Michigan
(www.umich.edu/$\sim$rotse). This research has made use of the SIMBAD
database, operated at CDS, Strasbourg, France, the STScI Digitized Sky
Survey (DSS) (http://archive.stsci.edu/cgi-bin/dss\_form), the International
Astronomical Union's ``List of Supernovae''
(http://cfa-www.harvard.edu/iau/lists/Supernovae.html), operated by the
Central Bureau for Astronomical Telegrams (CBAT), and the NASA/IPAC
Extragalactic Database (NED), which is operated by the Jet Propulsion
Laboratory, California Institute of Technology, under contract with the
National Aeronautics and Space Administration.

\clearpage
\begin{deluxetable}{llll}
\tablewidth{0pt}
\tablecaption{Upper Limits on LSB Galaxies\label{tab:limits}}
\tablehead{
\colhead{Parameter} & \colhead{$1\sigma$ confidence} & \colhead{95\%
confidence} & \colhead{Observed values}}

\startdata
SN Ia observed & $\le 1.84$ & $\le 3.00$ \\
SN Ia rate $(\hbox{SN} \,\, \hbox{Mpc}^{-3} \,\, \hbox{century}^{-1})$ & $\le
  2.02 \times 10^{-3}$ & $\le 3.29 \times 10^{-3}$ \\
${{\cal L}_B} \, (L_B \, \hbox{Mpc}^{-3})$ & $\le 1.55 \times 10^8$ & $\le 2.53
  \times 10^8$ & $(1.35 \pm 0.14) \times 10^8$ \tablenotemark{a} \\
$\Omega_b$ & $\le 0.0025$ & $\le 0.0040$ & 0.044 $\pm$
  0.004 \tablenotemark{b} \\
$\Omega_m$ & $\le 0.022$ & $\le 0.036$ & 0.27
  $\pm$ 0.04 \tablenotemark{b} \\
\enddata

\tablenotetext{a}{Mean HSB value.}
\tablenotetext{b}{Mean value for the Universe.}

\tablerefs{The mean HSB ${\cal L}_B$ value is from \citet{fuk04}.
  The $\Omega_b$ and $\Omega_m$ values
  are from WMAP \citep{ben03}.}

\end{deluxetable}
\clearpage

\clearpage
\clearpage

\clearpage
\clearpage

\clearpage
\clearpage

\clearpage
\clearpage

\end{document}